\documentstyle[prl,epsf,aps]{revtex}

\newcommand \be {\begin{equation}}
\newcommand \bea {\begin{eqnarray}}
\newcommand \ee {\end{equation}}
\newcommand \eea {\end{eqnarray}}

\newcommand \bi {\bibitem}
\newcommand \s {\sigma}

\makeatletter
\newcommand\erfc{\mathop{\operator@font erfc}\nolimits}
\makeatother
\input{epsf}
\begin{document}
%\show\mathrm
\twocolumn[\hsize\textwidth\columnwidth\hsize\csname@twocolumnfalse\endcsname
\draft       

\title{A general method to determine replica symmetry breaking transitions}
\author{E. Marinari, C. Naitza, F. Zuliani}
\address{Dipartimento di Fisica and INFN, Universita di Cagliari\\
Via Ospedale 72, 07100 Cagliari (Italy)\\
E-Mail: marinari@ca.infn.it,naitza@ca.infn.it,zuliani@ca.infn.it}
\author{G. Parisi}
\address{Dipartimento di Fisica and INFN, Universita di Roma 
{\it La Sapienza}\\
P. A. Moro 2, 00185 Roma (Italy)\\
E-Mail:  giorgio.parisi@roma1.infn.it}
\author{M. Picco}
\address{{\it LPTHE}\\
       \it  Universit\'e Pierre et Marie Curie, PARIS VI\\
       \it Universit\'e Denis Diderot, PARIS VII\\
        Boite 126, Tour 16, 1$^{\it er}$ \'etage, 4 place Jussieu\\
        F-75252 Paris CEDEX 05, FRANCE\\
E-Mail: picco@lpthe.jussieu.fr}
\author{F. Ritort}
\address{Departament de Fisica Fonamental, Facultat de Fisica\\
Universitat de Barcelona, Diagonal 647\\
08028 Barcelona (Spain).\\
E-Mail: ritort@ffn.ub.es}

\date{\today}
\maketitle

\begin{abstract}
We introduce a new parameter to investigate replica symmetry breaking
transitions using finite-size scaling methods. Based on
exact equalities initially derived by F. Guerra this parameter is a
direct check of the self-averaging character of the spin-glass order
parameter. This new parameter can be used to study
models with time reversal symmetry but its greatest interest concerns models
where this symmetry is absent. We apply the method to long-range
and short-range Ising spin glasses with and without magnetic field as
well as short-range multispin interaction spin glasses.
\end{abstract} 

\vfill
%\pacs{05.30.-d, 64.60.Cn, 64.70.Pf, 75.10. Nr}
%{\bf \hfill cond-mat/9502045}

\vfill
%\newpage

%\baselineskip 6mm
\twocolumn
\vskip.5pc]

\narrowtext
%1234567
The subject of replica symmetry breaking has become an important issue
in statistical physics \cite{MPV}. Since replica symmetry breaking was
proposed long time ago \cite{PARISI} there have been several new
developments concerning spin glasses as well as their applications in
other areas of statistical physics. The implications of the result
whether replica symmetry breaking takes place in real physical systems
can be of the outmost importance. Leaving aside the question whether and
how this transition could be observed in real experiments, it is
certainly relevant to establish the validity of mean-field theory for
spin glasses when applied to short-range systems. In this context, quite
recently a new controversy has appeared on the problem whether
self-averageness (i.e. the independence of the order parameter on the
microscopic realization of the quenched disorder) is automatically
satisfied in short-range systems. While the answer to this question
(according to heuristic arguments by Newman and Stein \cite{NS}) appears
to be closely related to the proper definition of the order parameter
and how the thermodynamic limit is taken, there are few doubts that
non self-averaging is the crucial signature for a spin-glass scenario where
replica symmetry breaks.

The purpose of this letter is to unambiguously show that indeed replica
symmetry (hereafter, referred to as RS) breaks in short-range spin-glasses
and that the genuine feature of the broken phase relies on the non-self
averaging character of the order parameter. While the major part of work
in spin glasses has been focused in models where there is a time reversal
symmetry in the Hamiltonian this is not an essential requirement for the
existence of a replica symmetry breaking (RSB) transition. In models
with time-reversal symmetry (hereafter referred as TRS), RS and TRS
break simultaneously at the spin-glass transition temperature. Because
both RS and TRS break precisely at the same temperature, it is very
difficult to distinguish the different features related to both
transitions. Indeed, the main distinction between the droplet
\cite{DROPLET} and the mean-field approaches relies on which symmetries
break at the spin-glass transition temperature. While in the first
approach only TRS breaks at the transition temperature, in the second
approach both symmetries break.  The most widely used parameter to
locate spin-glass transitions (the Binder parameter) signals the
breaking of time reversal symmetry rather than the other. Consequently,
the major part of numerical calculations using the Binder parameter do not
show that RS breaks at the spin-glass transition temperature but rather
whether TRS breaks. Then, it is essential to look for signatures of
replica symmetry breaking in models where time reversal symmetries are
lacking.

A large class of models where TRS is not present are
spin glasses in an external magnetic field or multispin p-interactions
spin-glass models (p-SG) with $p$ odd. The first class of models can be
described by Hamiltonians of the type,

\be
{\cal H}={\cal H}_0-h\sum_i\s_i=-\sum_{(i,j)}J_{ij}\s_i\s_j-h\sum_i\s_i
\label{eq1}
\ee

where the term $h\sum_i\s_i$ breaks the TRS ($\s\to-\s$) of the
Hamiltonian ${\cal H}_0$. On the other hand, models of p-SG take the
general form,

\be {\cal
H}=-\sum_{(i_1,i_2,..,i_p)}J_{i_1i_2..i_p}\s_{i_1}\s_{i_2}...\s_{i_p}
\label{eq2}
\ee

For $p$ odd TRS is absent. The interest of this last family of models
(contrarily to (\ref{eq1})) relies on the fact that there is no
parameter which appropriately tuned restores TRS (this happens in the
family of models of eq.(\ref{eq1}) where TRS is recovered if $h=0$).

When studying phase transitions in ordered systems one generally
computes the temperature dependence of cumulants of the order
parameter distribution such as susceptibilities (for instance, second
moments) which usually display power law divergences. Also, one can
consider adimensional parameters such as the kurtosis (usually known
as Binder parameter) or the skewness of the order parameter
distribution. These adimensional parameters are related to the
amplitudes of the scaling quantities in the renormalization group
flows and they are good indicators for the transition. They find their
most successful application in finite-size scaling studies.

The usefulness of these quantities to distinguish RSB transitions is
hampered by the fact that finite-size corrections to the leading
scaling behavior of the Binder parameter can be big. For RSB
transitions it is then convenient to consider adimensional quantities
which depend on other genuine features of the transition (and not only on
TRS) such as self-averageness. Our purpose here is to define a
suitable parameter which is the analogous of the Binder parameter for
transitions where TRS breaks and which can be used to locate
spin-glass transitions where RS breaks. In spin-glasses the order
parameter is not the global magnetization but a measure of the
freezing of the spins, the Edwards-Anderson parameter $q$
\cite{EA}. The appropriate way to compute this parameter is to consider
two replicas (i.e. two identical copies of the same system) and
compute the overlap, $q=(1/V)\sum_{i=1}^V\sigma_i\tau_i$ where $V$ is
the size or volume of the system. Expectation values of the moments
$\langle q^k\rangle_{BG}$ allow to reconstruct the order parameter
distribution $P_J(q)$ where $\langle ...\rangle_{BG}$ stands for the
usual Boltzmann-Gibbs average for a given sample.  It has been
recently shown by F. Guerra \cite{GUERRA} that sample to sample
fluctuations of the cumulants of the order parameter distribution
$P_J(q)$ are Gaussian distributed in the thermodynamic limit. For
instance, the following relationship is fulfilled in spin glasses in
the low temperature phase below $T_c$,

\be
G=\frac{\overline{\chi_{SG}^2}-\overline{\chi}_{SG}^2}
{\overline{V^2\langle(q-\langle q\rangle_{BG})^4\rangle_{BG}}-
\overline{\chi}_{SG}^2}=\frac{1}{3}
\label{eq4}
\ee

where $\overline{(.)}$ means average over the quenched disorder and
$\chi_{SG}$ (the spin-glass susceptibility) is
defined as

\be
\chi_{SG}=V(\langle q^2\rangle_{BG}-\langle q\rangle_{BG}^2)
\label{eq3}
\ee

The interest of defining the parameter $G$ is that it vanishes above the
transition temperature in the disordered phase where 
sample to sample fluctuations of $P_J(q)$ disappear in the $V\to\infty$ limit.
Similar information to that obtained from (\ref{eq4}) can be also 
gathered from the sample to sample fluctuations of $\chi_{SG}$,

\be
\alpha=\frac{\overline{\chi_{SG}^2}-\overline{\chi}_{SG}^2}
{\overline{\chi}_{SG}^2}
\label{eq5}
\ee

which only involves the first and second moments of the order
parameter. As we will see later (\ref{eq5}) yields also non trivial
behavior in the low temperature phase even though (in contrast to $G$)
it does not necessarily converge (in the thermodynamic limit) to a
temperature independent value. Consequently, $G$ is a parameter which
plays the same role as the usual Binder parameter $g$ in ferromagnets
and is given (in the $V\to\infty$ limit) by
$G(T)=(1/3)(1-\Theta_H(T-T_c))$ where $\Theta_H$ is the Heaviside
theta function. In RSB transitions (\ref{eq4}) goes to zero (as the
size $V$ increases) as $1/V$ for $T>T_c$ but converges to a finite
value for $T<T_c$. We expect the critical temperature (where RS
breaks) to be signaled by the crossing of the different curves
corresponding to different lattice sizes. Furthermore, close to $T_c$
it is reasonable to expect, $G(T)\sim \hat{G}(L/\xi)$ where $\xi$ is a
correlation length.  We stress that the calculation of $G$ is
specially useful in models where $TRS$ is absent. In the presence of
$TRS$ the usual Binder parameter $g$ can be used to locate the phase
transition with much less numerical effort. But the interest of $G$ is
that it emphasizes the non-selfaveraging character of the low
temperature phase.

%In
%analogy to what happens in second order phase transitions (where
%Boltzmann-Gibbs cumulants of the magnetisation are related to amplitudes
%of the renormalization group flows around the fixed point), now we
%expect (for RSB transitions) that disorder cumulants of non-self
%averaging quantities such as the ``Boltzmann-Gibbs cumulants of the
%order parameter'' (for instance, the spin-glass susceptibility) are the
%quantities around which the renormalization groups flow become
%meaningful. 

%Close to $T_c$ we could
%expect, $\alpha(T)\sim \hat{\alpha}(L/\hat{\xi})$. {\it A priori} the
%correlation length $\hat{\xi}$ could be different from the usual
%correlation length associated to the divergence of $\chi_{SG}$
%\cite{FOT2}. Consequently, we expect the critical temperature (where RS
%breaks) to be indicated by the crossing of the different curves
%corresponding to different lattice sizes. The behavior should be similar
%to what is found for usual cumulants of the magnetisation (such as the
%Binder parameter) in ferromagnets.

To check these predictions we have performed numerical simulation of
models of the previous type (\ref{eq1}) (with and without magnetic
field) and the $p$-spin model (\ref{eq2}). All the simulations use the
parallel tempering method, an efficient algorithm to thermalize small
samples \cite{TEMP}.  We have studied three different models, the
mean-field Sherrington-Kirkpatrick (SK) model \cite{SK}, the four
dimensional (4D) Ising spin glass, an
the model eq.(\ref{eq2}) in four dimensions with $p=3$. In this last
case, the Hamiltonian is short-ranged, there are two spins per site in a
4D simple cubic lattice and the Hamiltonian couples all possible
triplets of spins occupying nearest-neighbor sites of the lattice. More
precisely the Hamiltonian reads,

\bea {\cal
H}=-\sum_{i=1}^V\sum_{\mu=1}^D(J^{i,\mu}_{(12,1)}\s_1^i\s_2^i\s_1^{i+\mu}+J^{i,\mu}_{(12,2)}\s_1^i\s_2^i\s_2^{i+\mu}
+\nonumber\\J^{i,\mu}_{(1,12)}\s_1^i\s_1^{i+\mu}\s_2^{i+\mu}+
J^{i,\mu}_{(2,12)}\s_2^i\s_1^{i+\mu}\s_2^{i+\mu})
\label{eq6}
\eea

where the pair $(i,\mu)$ denotes the link of the lattice and the spins 
$(\s_1^i,\s_2^i)$ occupy the same site $i$ in the lattice. 

\begin{figure}
\begin{center}
\leavevmode
\epsfysize=230pt{\epsffile{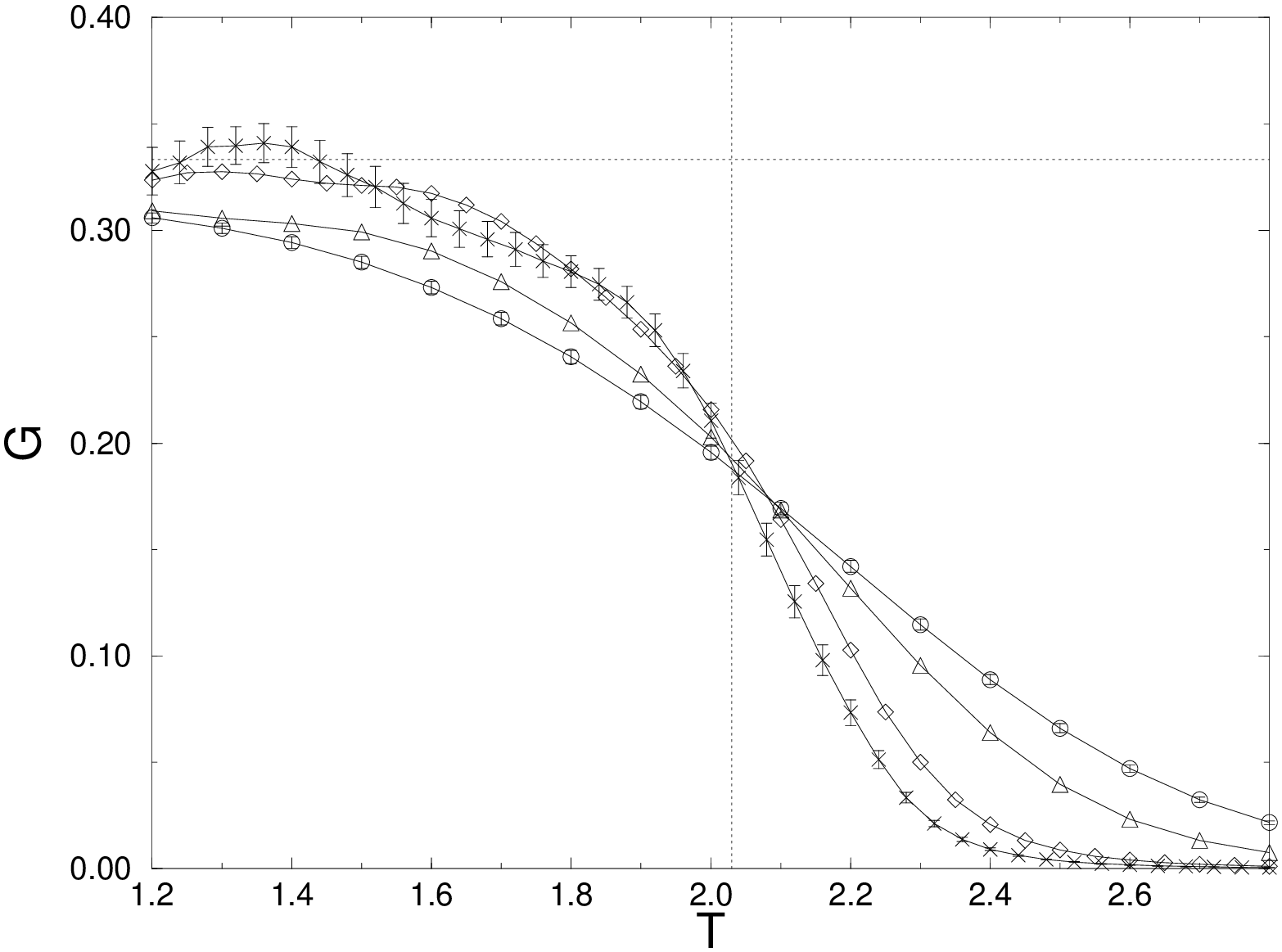}}
\end{center}
  \protect\caption[2]{$G$ in the 4D Ising spin glass without a field.
The horizontal line indicates the expected low $T$ result $G=1/3$ while
the vertical line indicates the expected transition temperature derived
form other methods \cite{HT,BCPRPR}. Error bars are shown for
$L=4,10$.\protect\label{FIG1} }
\end{figure}

First, we show the results in the four dimensional Ising spin glass
without a magnetic field ($h$=0). This is a check of our method since
the transition is well known using standard methods \cite{TEMP}. The
model is described by eq.(\ref{eq1}) with the $J_{ij}=\pm 1$
connecting nearest-neighbor sites of in a cubic lattice of side $L$
with periodic boundary conditions. The simulations were done for sizes
$L=4,5,8,10$ ($2944,1920,1376,320$ samples respectively) with $100000$
Monte Carlo steps (MCS) of thermalization time and the same amount of steps
to collect statistics (for $L=10$ we did runs up to 35 million of
MCS). Figure 1 shows the results for $G$. Note the
existence of a critical temperature above which $G$ goes to zero and
below which it converges to 1/3. The different curves cross at a
temperature in agreement with that derived from the analysis of the
usual Binder parameter \cite{BCPRPR} and also series expansions
\cite{HT} ($T_c\simeq 2.03$). We have also analyzed the parameter
$\alpha$ of eq.(\ref{eq5}) which shows that $\alpha(L)\sim 1/V$ for
$T>T_c$ while $\alpha(L)\simeq \alpha(\infty)+A/L^{\lambda}$ with
$A>0$ for $T<T_c$ giving qualitatively the same information as $G$.

Next we consider models without TRS. We first consider the study of
the SK model in a magnetic field. The SK model \cite{SK} corresponds
to eq.(\ref{eq1}) with $J_{ij}$ long-ranged and Gaussian distributed
with $\overline{J_{ij}}=0, \overline{J_{ij}^2}=1/V$. The existence of
a transition in a field is well established in mean-field theory but
there are few results which corroborate its existence using numerical
simulations \cite{PR}. Figure 2 shows $G(T)$ for $V=32,256,512,1024$
with 1000,1000,400,150 samples respectively.  While it is very
difficult to see evidence for this transition with the usual cumulants
(skewness or Binder parameter) the situation turns out to be more
clear with the parameter $G$ where a merging close to $T\simeq .6-.7$
(below $T_c(h=0)=1$) is observed. The figure clearly shows the
existence of two temperature regions. A high temperature region where
$G$ goes to zero with the volume (as $1/V$) and a low temperature one
where $G$ converges to 1/3 (within the precision of the
statistics). This shows the existence of the Almeida-Thouless line in
the SK model, a result well known in the mean-field theory of spin
glasses but difficult to observe numerically.

\begin{figure}
\begin{center}
\leavevmode
\epsfysize=230pt{\epsffile{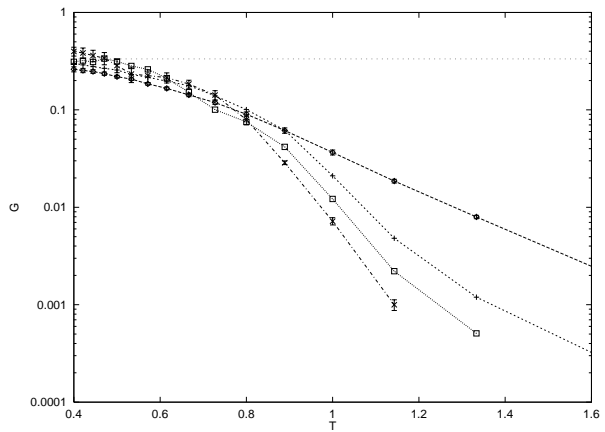}}
\end{center}
  \protect\caption[1]{$G$ in the SK model at $h=0.3$. Error bars are shown
  for $V=32,1024$. The different curves merge at a temperature well
  compatible with the theoretical result $T_c(h=0.3)=0.65$
  \cite{AT}.\protect\label{FIG2} }
\end{figure}

The results in the four dimensional Ising spin-glass model in a field
are shown in figure 3. Simulations were done at a magnetic field $h=0.4$
with statistics ranging from 20000 MCS for $L=3$ up to $450000$ for
$L=9$. We observe also here a behavior very similar to that found in
figure 2. The existence of the two regions (a high temperature one where
$G$ goes to zero and a low temperature one where $G$ converges to a
finite value close to 1/3) is also clear from the plot.

Figures 2 and 3 show quite unambiguously that there are two regions
where self-averaging properties are quite different. This is a strong
indication in favor of the existence of a RSB phase transition in spin
glasses in a magnetic field. But a scale invariant crossing point is not
so clearly observed in figures 2 and 3 compared to what is observed in
figure 1 for zero magnetic field and figure 4 (see below). There are two
factors which make numerical simulations of spin glasses in a magnetic
field much more difficult. The first one is related to the difficulty of
thermalizing spin-glasses at low temperatures. When the field is switched
on the critical temperature is pushed down (as figures 2 and 3 clearly
show). This makes thermalization in the low temperature region
more difficult. The second factor is related to the fact that
Eq.(\ref{eq1}) restores the TRS at zero magnetic field. Consequently, it
is natural to expect the existence of a crossover length $L_c$ (which
increases as the magnetic field decreases) such that above $L_c$ the
finite-field fixed point dominates the scaling behavior while below
$L_c$ the scaling behavior is dominated by the zero field fixed
point. In the case of figure 3 the crossover length was previously
estimated ($L_c\simeq 5$ \cite{PR}). This crossover effect manifests as
a displacement of the crossing point to lower temperatures as the size
increases. For large sizes (and always within errors) the crossing point
stays around $T_c\simeq 1.2$, a value for the critical temperature which
has been estimated also through other methods \cite{PR2,MPZ}.

\begin{figure}
\begin{center}
\leavevmode
\epsfysize=230pt{\epsffile{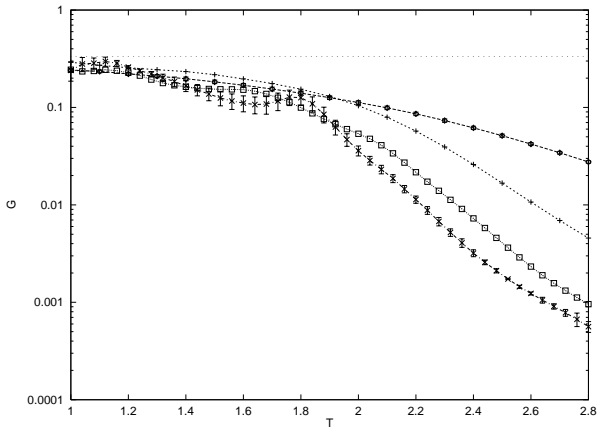}}
\end{center}
  \protect\caption[1]{$G$ in the 4D $\pm J$ Ising spin glass at
  h=0.4. The number of samples is $2560,1280,704,64$ for $L=3,5,7,9$
  respectively. Error bars are shown for $L=3,9$.\protect\label{FIG3} }
\end{figure}

Assuming that the value of the parameter $G$ at the crossing point
corresponds to a universal amplitude we find (after examination of the
data for the SK and the 4D Ising spin glass at zero and finite field),
that $G_c=G(T=T_c)$ clearly increases with the field. This result
suggests (in case the previous assumption is correct) that the
transition without a field and in a field are determined by different
fixed points \cite{BR}.

\begin{figure}
\begin{center}
\leavevmode
\epsfysize=230pt{\epsffile{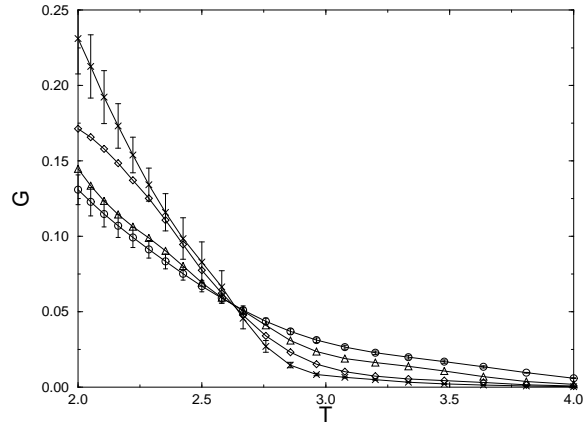}}
\end{center}
  \protect\caption[4]{$G$ in the model (\ref{eq6}) without TRS symmetry
with three-spins interaction and two spins per site. We find that
$T_c\simeq 2.62$. Error bars are shown for $L=3,6$.\protect\label{FIG4}
}
\end{figure}

To check that the parameter $G$ is indeed a good tool to determine RSB
transitions it would be more convenient to consider a model where
there is no external small parameter (like the field) which can
restore the TRS. For such a model there will not be a crossover length
$L_c$ and a crossing point for the parameter $G$ should be easier to
see already for small sizes. To confirm these expectations we have
investigated model (\ref{eq6}) with $J^{i,\mu}=\pm 1$ in 
four dimensions in lattices of sizes
$L=3,4,5,6$ with $100000$ MCS of statistics per
temperature. The results are shown in figure 4.  From (\ref{eq6}) it
is clear that there is no global symmetry in this model and the only
possible phase transition we can expect is a first-order one. We have
verified that there is no latent heat and that the order parameter
$\langle q\rangle_{BG}$ does not experience any discontinuous jump at
any finite temperature. Having discarded the usual thermodynamic
first-order transition scenario the only possibility left is the
existence of a RSB transition where the spin-glass susceptibility
diverges. Indeed our results show an algebraic divergence of the
spin-glass susceptibility $\chi_{SG}$ and a least-squares fit gives
$\chi_{SG}\sim (T-T_c)^{-\gamma}$ with $T_c\simeq 2.63$ and
$\gamma\simeq 1.0$. This value of $T_c$ is in striking agreement with
the crossing point observed in figure 4. If one assumes, as said
before, that $G(T)\sim \hat{G}(L/\xi)$ with $\xi\sim (T-T_c)^{-\nu}$
then $(dG/dT)_{T=T_c}\sim L^{\frac{1}{\nu}}$. A power law fit yields
$\nu\simeq 1/2$ suggesting that both $\gamma$ and $\nu$ are close to
mean-field values. Let us note that the same conclusions are obtained
by studying the parameter $\alpha$ in eq.(\ref{eq5}).

Summarizing, we have proposed a new parameter $G$ based on exact
inequalities initially derived by Guerra \cite{GUERRA}. This parameter
is suited to numerically study replica symmetry breaking
transitions. The physical meaning of $G$ is related to the very nature
of the replica symmetry broken phase, i.e. the absence of
self-averaging in the order parameter. When replica symmetry is not
broken the spin-glass order parameter is self-averaging but this
property is lost in the broken phase. The parameter $G$ in
eq.(\ref{eq4}) has the good properties of being bounded and positive
(a property which does not have the usual Binder parameter used for
spin-glasses without TRS) and can be used as a good indicator for RSB
transitions using finite-size scaling methods. At high temperatures
$G$ goes to zero as $1/V$ where $V$ is the volume of the system while
at low temperatures it converges to 1/3 in the $L\to\infty$ limit. The
method can be also used to investigate transitions where TRS
breaks. In particular, we have investigated the 4D Ising spin-glass
at zero field and found that indeed RS breaks at $T_c$.  This result
unambiguously shows that both TRS and RS break at $T_c$. But its
greatest potential applicability concerns disordered models without
TRS. In particular, we have considered spin glasses in a magnetic
field (the SK model as well as the 4D Ising spin glass) which clearly
show the existence of two different regimes corresponding to different
self-averaging properties of the order parameter. But the nicest
application of the method is for models where there is no tunable
parameter which restores time reversal symmetry (like the magnetic
field). By introducing a new short-range $p$-spin model
(eq.(\ref{eq6})) we have shown that $G$ is indeed a good indicator for
RSB. We have considered a model with $p=3$ in 4D and we have shown
that there is no thermodynamic first phase order transition. In this
case $G$ displays a crossing point where the spin-glass susceptibility
diverges. Finally, we want to stress that the information gathered
from $G$ in models without TRS cannot be extracted in an easy way from
the usual standard cumulants of the sample averaged $P(q)$. The
genuine property of replica symmetry breaking transitions in
disordered systems is the non self-averaging character of the
spin-glass order parameter, a feature which is specifically taken into
account within the present method. A more deep understanding of the
appropriate renormalization group approach in spin glasses is certainly
needed to clarify the appropriate theoretical framework to deal with
this type of phase transitions.

{\bf Acknowledgments}. F.R acknowledges Theo Nieuwenhuizen for 
discussions and a careful reading of the manuscript. 

\hspace{-2cm}

\vfill

\end{document}